\documentclass[twocolumn,aps,superscriptaddress,showpacs,nofootinbib,floatfix]{revtex4}
\usepackage{listings}
\usepackage[utf8]{inputenc}
\usepackage{hyperref}
\usepackage{amsmath}
\usepackage{amssymb}
\usepackage{amsfonts}
\usepackage{tabularx}
\usepackage{booktabs}
\usepackage{graphicx}
\usepackage{color}
\usepackage{multirow}
\usepackage{setspace}
\usepackage[inline]{enumitem}
\usepackage{makecell,rotating,diagbox}
\usepackage[T2A,T1]{fontenc}
\graphicspath{{fig/}}
\definecolor{theblue}{RGB}{0,50,230}

\usepackage{hyperref}
\hypersetup{
  colorlinks=true,
  linkcolor=theblue,
  citecolor=theblue,
  urlcolor=theblue
}

\newlength\cmsFigWidth
\usepackage[normalem]{ulem}  

\renewcommand\sout{\bgroup \color{red} \ULdepth=-.5ex \ULset}

\begin{document}

\title{Investigating student understanding of heat engine: a case study of Stirling engine}

\author{Lilin Zhu}

\author{Gang Xiang}\email{gxiang@scu.edu.cn}
\affiliation{Department of Physics, Sichuan University, Chengdu 610064, China}

\date{\today}

\begin{abstract}
We report on the study of student difficulties regarding heat engine in the context of Stirling cycle within upper-division undergraduate thermal physics course. An in-class test about a Stirling engine with a regenerator was taken by three classes, and the students were asked to perform one of the most basic activities---calculate the efficiency of the heat engine. Our data suggest that quite a few students have not developed a robust conceptual understanding of basic engineering knowledge of the heat engine, including the function of the regenerator and the influence of piston movements on the heat and work involved in the engine. Most notably, although the science error ratios of the three classes were similar ($\sim$10\%), the engineering error ratios of the three classes were high (above 50\%), and the class that was given a simple tutorial of engineering knowledge of heat engine exhibited significantly smaller engineering error ratio by about 20\% than the other two classes. In addition, both the written answers and post-test interviews show that most of the students can only associate Carnot's theorem with Carnot cycle, but not with other reversible cycles working between two heat reservoirs, probably because no enough cycles except Carnot cycle were covered in the traditional Thermodynamics textbook. Our results suggest that both scientific and engineering knowledge are important and should be included in instructional approaches, especially in the Thermodynamics course taught in the countries and regions with a tradition of not paying much attention to experimental education or engineering training.
\end{abstract}
\maketitle

\section{Introduction}\label{sec:intro}
Any device that transforms heat partly into work or mechanical energy is called a heat engine. All engines absorb heat from a source at a relatively high temperature, perform some mechanical work, and reject some heat at a lower temperature. About two hundred years ago, a French engineer Sadi Carnot developed a hypothetical, idealized heat engine that has the maximum possible efficiency consistent with the second law of thermodynamics \cite{Klein:1974pt}. The cycle of the engine is called Carnot cycle. Almost at the same time (8 years earlier), another cycle called Stirling cycle was proposed by Robert Stirling, which is known as a reversible cycle as basic and important as Carnot cycle in thermodynamics \cite{Spence:1982pt, Crane:1990pt, Maclsaac:2004pt}. In fact, Stirling engine was one of the earliest heat engines put to practical use and is still used in various facilities such as submarines and concentrated solar power system. Other basic cycles such as Otto cycle \cite {Mozurkewich:1982jap} and Diesel cycle \cite{Hoffmann:1985jap} were proposed later in 19th century and have been used in the engines of various vehicles such as cars since then.

\begin{figure}[pht]
  \includegraphics[scale=0.4]{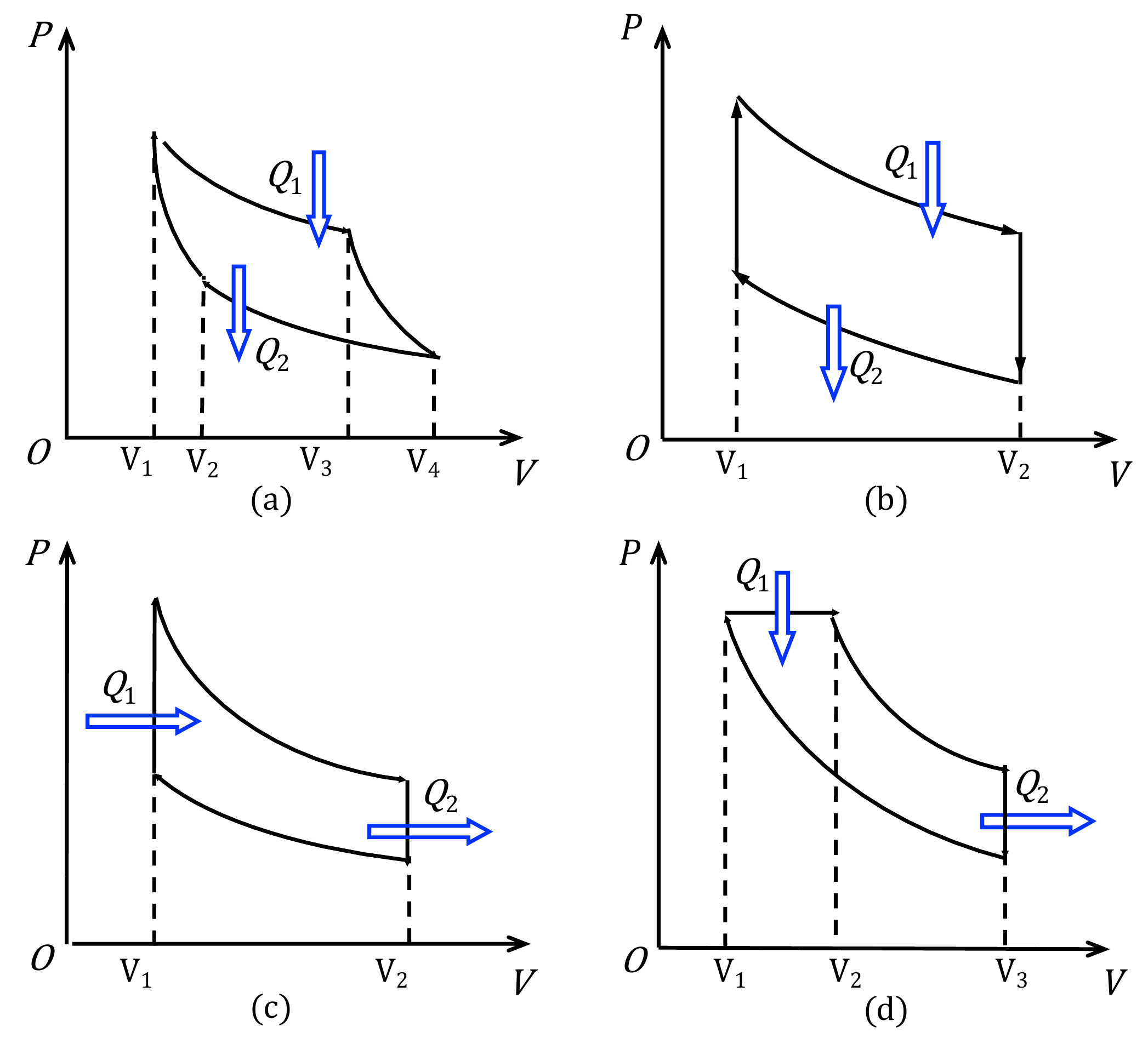}
  \caption{The P-V diagrams for (a) Carnot cycle, (b) Stirling cycle, (c) Otto cycle and (d) Diesel cycle. $Q_1$ and $Q_2$ represent the quantities of heat absorbed and rejected by the engine during one cycle, respectively.}
  \label{fig1}
\end{figure}

Fig. \ref{fig1} shows four classical cycles: (a) Carnot cycle, (b) Stirling cycle, (c) Otto cycle and (d) Diesel cycle. Among them, both Carnot cycle and Stirling cycle are used in external combustion engines, while Otto cycle and Diesel cycle are used in internal combustion engines. Interestingly, all the Thermodynamics textbooks describe Carnot cycle in great detail, but many of them just introduce the other three cycles briefly \cite {Young:2010} and some of them even don't cover them at all. Coincidentally, or perhaps correspondingly, a number of studies in physics education research have shed light on student understanding of basic concepts of Carnot cycle \cite{Leff:2018ajp,Cochran:2006ajp,Smith:2009acp,Smith:2015prs, Meltzer:2005aip}, but very few studies have focused on the topics related to Stirling cycle, Otto cycle or Diesel cycle \cite{Simon:1984ajp, Shieh:2014ajp, Romanellia:2017ajp, Romanelli:2020ajp}, which are widely used in automobiles and many other types of machinery. In this sense, evaluating of student understanding of basic knowledge of different kinds of heat engines, for instance, Carnot engine and Stirling engine, is important and necessary for the development of the pedagogical content knowledge (PCK) needed to teach the heat engine well, not only in physics, but also in related applied disciplines. It is worthwhile noting that both Carnot cycle and Stirling cycle are reversible cycles working between two heat reservoirs, which means Carnot's theorem are applicable to both of them.

Among various educational evaluation techniques, post-testing is the most widely adopted method in the physics education research (PER) \cite{Fraenkel:1996, Sundberg: 2002cbe, Bao:2006prs}. PER focuses on understanding how students learn physics at all levels and developing strategies to help students learn physics more effectively \cite{Singh:2014pt, Stith:2002ajp}. Based on these facts, we designed an in-class test about Stirling cycle at the end of the course and then performed this study. The objectives of this study are as follows. First of all, Stirling cycle is a good example for the students to understand how to calculate the efficiency of the cycle, since it involves a regenerator that alternatively absorbs and releases heat, which is unique (and puzzling) and not found in any other cycles including Carnot cycle. Secondly, according to Carnot's theorem, every reversible heat engine between a pair of heat reservoirs, such as Carnot engine and Stirling engine, is equally efficient, regardless of the working substance or the operation details. However, it is possible that some students would take it for granted that the heat engine here only refers to Carnot engine if other types of heat engines are not taught or just introduced briefly. In this sense, our study can estimate to what extent the students understand Carnot's theorem. Finally, it is noted that in the last decade or so, a growing number of researchers have shown interest in student understanding of thermodynamics beyond the theoretical level, leading to upper-division engineering instruction on heat engines \cite{Bejan:1994ajp, Clark:2013aip, Meltzer:2007aip}.Interestingly, Stirling engine with a regenerator in this study is such an exquisite design that the more the students work on it, the more the students will be stimulated to be curious in thermodynamics and be interested in integrating theory with practice.

This paper is organized as the following: we first introduce Stirling cycle in Sec. II. A brief overview of the research context and methodology is then presented in Sec. III. Then we discuss the student difficulties identified as well as implications for instruction. Finally, a summary is given in Sec. V.

\begin{figure*}[pht]
  \includegraphics[scale=0.45]{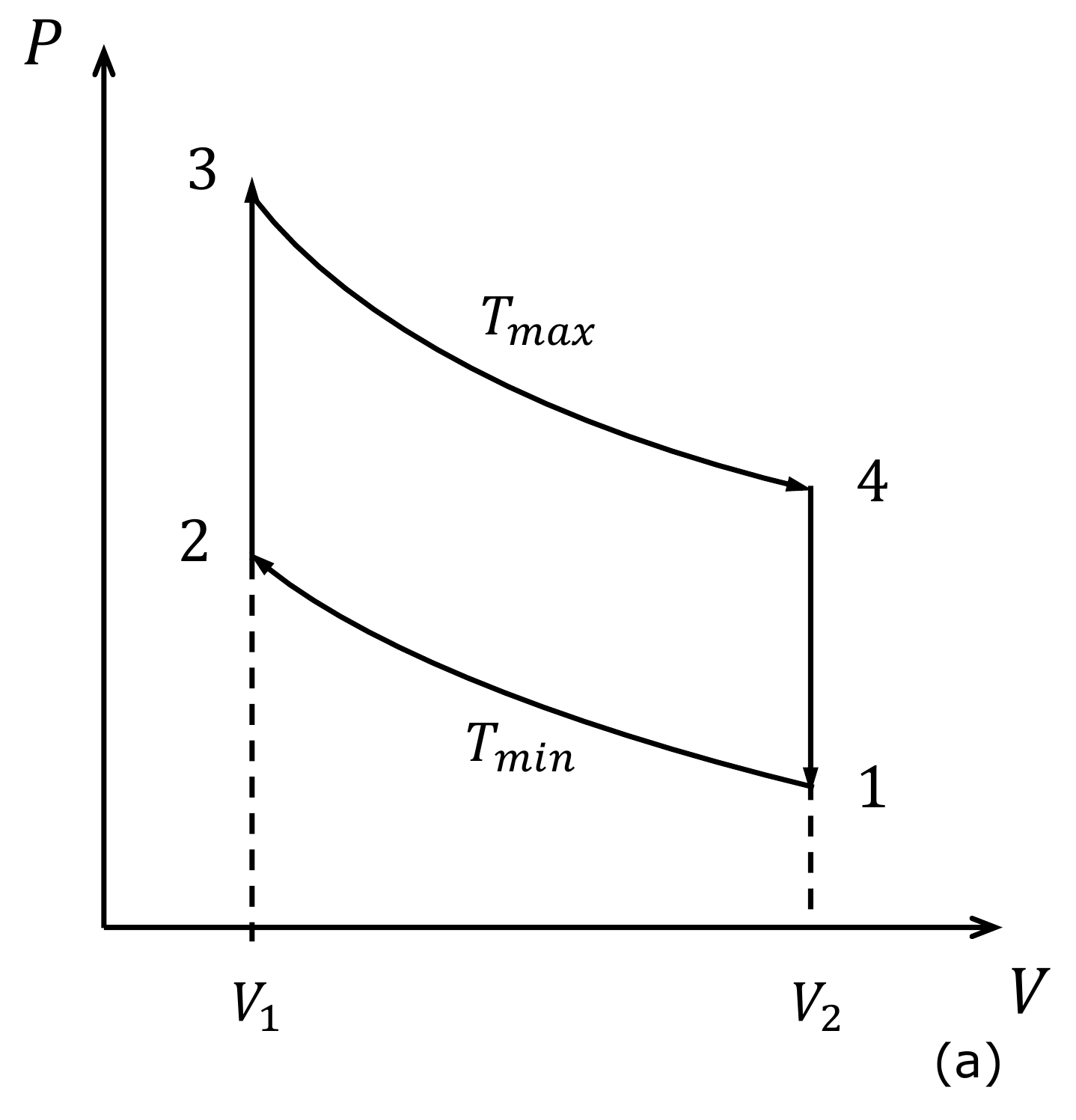}
  \hspace{1cm}
    \includegraphics[scale=0.35]{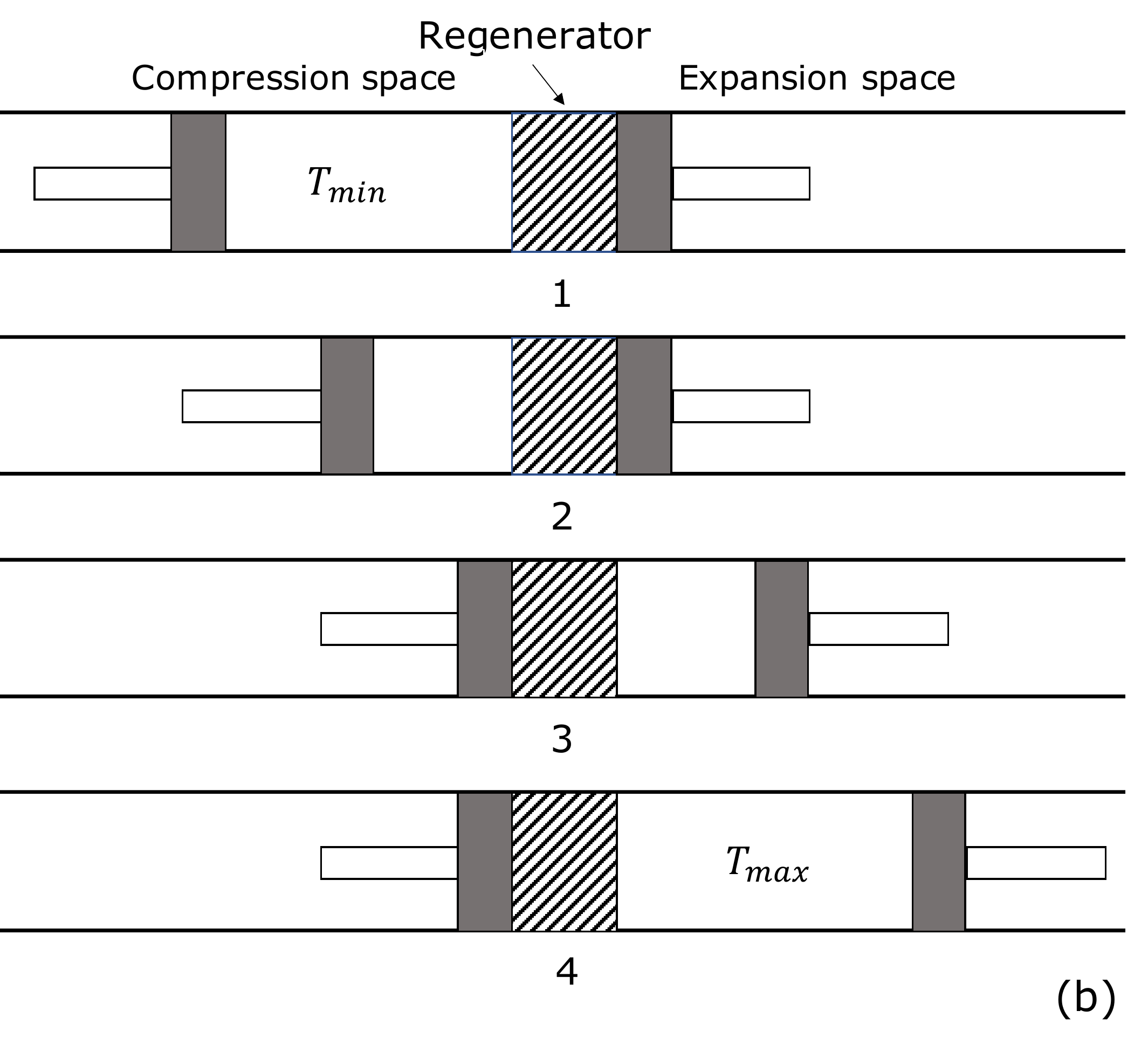}
   \begin{itemize}
  \item [*] {\bf\underline {Question}}: {\it The Stirling cycle is shown above: (a) P-V diagram; (b) piston arrangement at the four terminal points of the cycle. Assuming that the working material is an ideal diatomic gas and the pistons move without friction or leakage of the gas, calculate the efficiency of the heat engine. } 
  \end{itemize}
  \caption{Question used in this study. Administered after lecture instruction.}
  \label{fig2}
\end{figure*}

\section{Stirling cycle}
A Stirling cycle is similar to a Carnot cycle, except that the two adiabatic processes are replaced by two constant-volume processes. As shown in Fig. \ref{fig2}(a), a Stirling cycle involves two isothermal processes $1\to 2$ and $3\to 4$, as well as two constant-volume processes $2\to 3$ and $4\to 1$. In practice, a Stirling heat engine includes a cylinder containing two opposed pistons and a regenerator between the pistons. The regenerator serves as a thermodynamic sponge that alternatively absorbs and releases heat. The cylinder is divided by the regenerator into two parts: the volume between the regenerator and a piston maintained at high temperature $T_{max}$ is called the expansion space, while the other, maintained at low temperature $T_{min}$ the compression space, as illustrated in Fig. \ref{fig2}(b), which shows the piston arrangement at the terminal points of Stirling cycle. At the beginning, all the working gas is in the compression space with the temperature $T_{min}$, which is corresponding to the point $1$ in Fig. \ref{fig2}(a). In the constant-volume process $2\to 3$, the working gas is transferred from the compression space to the expansion one through the regenerator, which is heated from $T_{min}$ to $T_{max}$. During the final process $4\to 1$, both pistons move simultaneously, transferring the working gas from the expansion space to the compression one. In the passage through the regenerator, the working gas releases heat and emerges into the compression space at $T_{min}$.

The question on Stirling cycle, shown in Fig. \ref{fig2}, was developed to assess student understanding of the efficiency calculation of heat engine. It should be emphasized that during the cycle, the heat transferred in the constant-volume processes $2\to 3$ and $4\to 1$ is contained in the regenerator. Therefore, the regenerator adds an additional layer of complexity. It is pedagogically useful, since it can be used to assess or facilitate deeper conceptual understanding of important quantitative and qualitative properties of heat engines.

\section{methods}\label{sec:methods}
This investigation was carried out with three classes of physics undergraduates enrolled in Thermodynamics and Statistical Mechanics course at Sichuan University (SCU) in 2019, which is a renowned four-year public research university in China. This course is typically taken during the first semester of the junior year. The sizes of the classes varied from 14 to 54 students. The students were given course credit for participating in in-class tests during the semester and a final exam at the end. The three classes were taught by different instructors, but the lectures, homework assignments and in-class tests were presented in traditional formats. The only nontraditional aspect was the addition of voluntary tutorial given by one of the authors (Xiang), which emphasized not only the theoretical (science) aspects, but also the experimental (engineering) aspects of the basic knowledge of thermodynamics, for instance, how the refrigerator is scientifically and mechanically related to Carnot cycle and how the engineering improvements of internal combustion engines have influenced the scientific and technological advances of our society and impacted our daily life. This tutorial is new and unusual to the participants, especially when one recognizes that traditional college courses of physics in China do not attach much importance to experimental education or engineering training. It is noted that neither the structure of Stirling engine, nor the calculation of the efficiency of Stirling cycle was covered in the tutorial. Therefore, the preliminary analysis showed no significant differences in answering the questions of what Stirling cycle is and how it works between tutorial participants and nonparticipants.  

This work focuses on the results from a particular question from the in-class test, shown in Fig. \ref{fig2}. The students completed the test individually in quiet rooms and were cut off after 30 minutes. In this study, the working material is assumed to be an ideal diatomic gas for the calculation of the cycle efficiency. The heat transferred in the constant-volume process $2\to 3$ is equal to that in the process $4\to 1$ for ideal Stirling cycle. In this case, one only need to consider the heat supply in the process $3\to 4$ and the heat rejection in the process $1\to 2$ between the engine and its surroundings. This heat supply and heat rejection at constant temperature satisfies the requirement of the second law of thermodynamics for maximum thermal efficiency, so that the efficiency of Stirling cycle is the same as Carnot cycle. Any answer that satisfies a few key criteria was considered as a correct one. These criteria include correct calculations of heat rejection in the isothermal process $1\to 2$, heat supply in the isothermal process $3\to 4$ and efficiency $\eta$ of the cycle, or clear expressions of using Carnot's Theorem to obtain correct efficiency for Stirling cycle.

Therefore, Stirling cycle not only gives us opportunity to estimate student understanding of general knowledge about the heat, work and efficiency involved in a heat engine, but also provides novelty to the students which can be used to explore their understanding of Carnot's theorem. Although the students have learned that all reversible heat engines between two heat reservoirs have the same efficiency, regardless of other details, our study will show that most of the students can only associate Carnot's theorem with Carnot cycle, but not with Stirling cycle. 

In addition to the collection and analysis of the written answers by the students, post-test interviews were also carried out with the students. Each interview lasted an average of 10 minutes. Instead of being constrained to cover all the questions about Stirling cycle, the interviews focused on the most commonly discussed or typical questions that were related to the student errors in the in-class test. During the interviews, the students were required to talk aloud, and the interviewers provided minimal interventions only for reminding the students to keep talking or asking them to clarify explanations not understood by the interviewers.

\section{results and analysis}\label{sec:dis}

The data from the in-class test on Stirling cycle were collected and analyzed at SCU in 2019 to investigate student performance. In the following, we discuss the test results in terms of qualitative and quantitative description.

\subsection{Qualitative description of student difficulties}
To better understand student thinking behind their written answers, interviews were conducted, which involved an undergraduate student and an interviewer each time. It was found that the most common error made by the students was the misunderstanding in the function of regenerator. As a result, the heat transferred in the two constant-volume processes $2 \to 3$ and $4 \to 1$ was included when calculating the efficiency of Stirling cycle. Some typical interviews are listed as follows.\\
{\bf \textit {Interview 1}}

I: Can you tell me why you consider the heat supply in the process $2 \to 3$ and heat rejection in the process $4 \to 1$ when calculating the efficiency of Stirling cycle?

S1: That's just how I learned. The heat supply and heat rejection of all processes should be included according to the formula of heat engine's efficiency.

I: You mean the heat transferred in the engine should also be considered ?

S1: Yes.

I: How about the regenerator? You didn't consider it during the calculation.

S1: Um... I'm not sure. Maybe it's designed for improving the efficiency.\\
{\bf \textit {Interview 2}}
 
I: Can you explain why you only include the heat supply and the heat rejection in the two isothermal processes $1\to 2$ and $3 \to 4$ when calculating the efficiency?

S2: Because in this way, I can get the efficiency of $(T_1-T_2)/T_1$.

I: You mean Stirling engine has the same efficiency as Carnot engine?

S2: Yes.

I: Okay, why?

S2: Um... I'm not sure.

I: Let me ask in a slightly different way. Where do the two constant-volume processes $2 \to 3$ and $4 \to 1$ happen?

S2: In the engine?

I: So you think the heat transferred for the whole cycle is independent of the regenerator?

S2: Yes. Actually, I think it is just for the practical engineering requirement.

Both S1 and S2 (and indeed most students interviewed) did clearly know that the efficiency of heat engine is defined as
\begin{eqnarray}
\eta=\frac{Q_{H}-Q_{C}}{Q_{H}}.
\label{efficiency}
\end{eqnarray}

This appeared to be extremely stable in that students never questioned this, even when confronted with other issues. It should be pointed out that $Q_{H}$ and $Q_{C}$ \footnote{The symbols are used to represent to absolute values of the energy transfers throughout a heat engine cycle. Therefore, they are inherently positive.} in Eq. (\ref{efficiency}) represent the quantities of heat transferred from the hot and cold reservoirs during one cycle, respectively. In other words, the heat transferred per cycle in the engine should not be contained. But most students misunderstood the physical meaning of $Q_{H}$ and $Q_{C}$. S1 was firmer in the belief that all heat transferred during the whole cycle should be considered. Actually, S1 didn't understand the physical meaning of efficiency of heat engine . S2 was trying to speak that all reversible engines have the same efficiency, but was unable to properly verbalize it. Even though Carnot's theorem is fundamental in the course of Thermodynamics, S2 was not familiar with it. Therefore, S2 also may not understand that only the heat transferred between the heat engine and hot (cold) reservoir rather than the total heat transferred during the whole cycle should be considered when calculating the efficiency of heat engine.

These interviews were fairly representative of the students as a whole. Most of the students were not able to describe how the regenerator works, indicating a lack of understanding of the relevant engineering concepts of heat engine. As indicated by the interviews, the students in general had a more varied mastery of the function of regenerator. In some cases, the students only treated the regenerator as a required engineering component without knowing its function. This appears to stem from the students' lack of engineering knowledge of heat engines. In the Thermodynamics course, students majoring in physics usually only learn the theoretical knowledge of heat engines. Relevant engineering aspects are not introduced in the classes, unless the instructor places emphasis on them. This assertion is supported by students' preformation in the test. Only three students gave correct description of the regenerator, who were from the same class. Their instructor (Xiang) did give them some general engineering introduction of heat engine.

\begin{figure*}[pht]
  \includegraphics[scale=0.5]{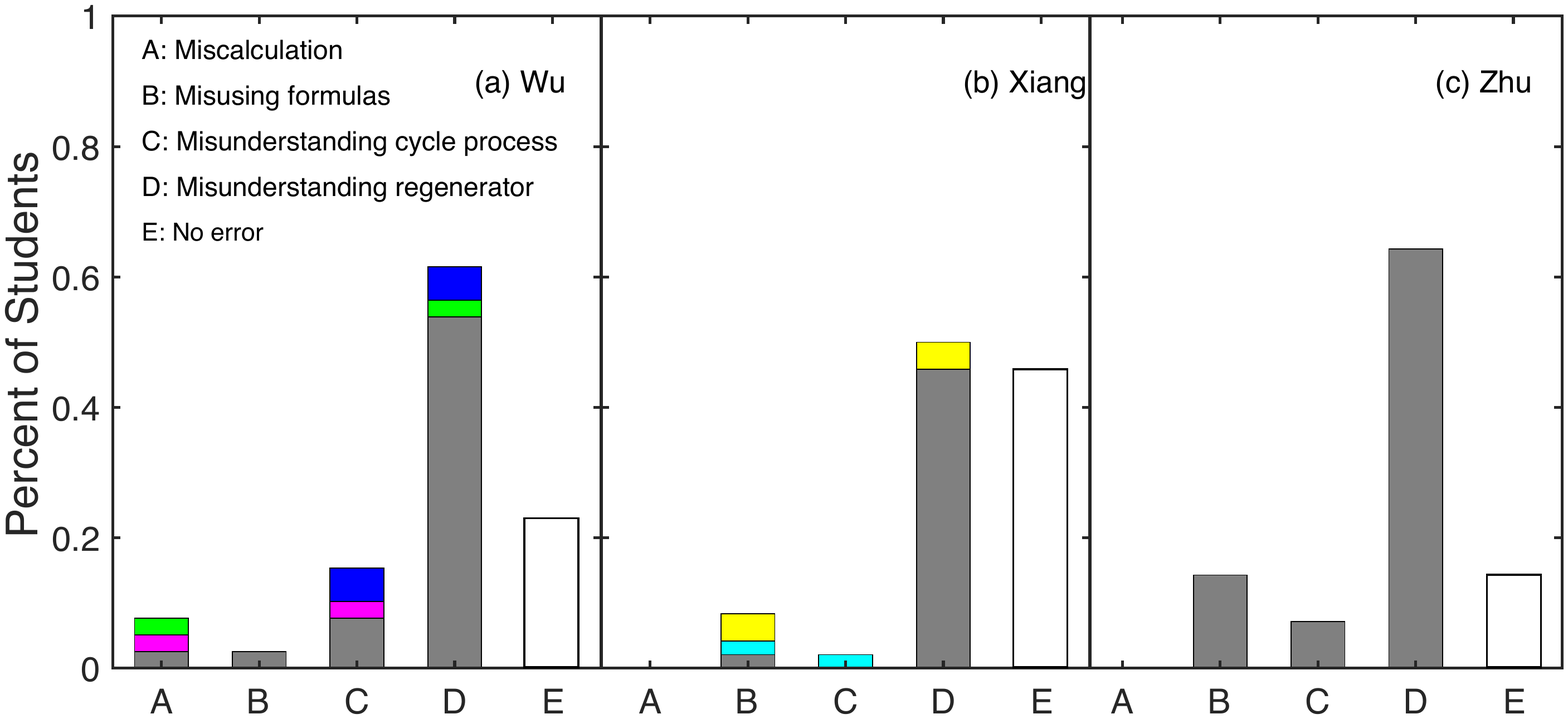}
  \vspace{-2cm}
  \caption{Comparison of student errors in the test on Stirling cycle for (a) Wu class, (b) Xiang class and (c) Zhu class. Some bars are composed of several parts with different colors. The grey part shows the percentage of students who only made one error, while the others represent some students made multiple errors, simultaneously. The case of no error is established by empty bar. The detailed explanation can be found in the context.}
  \label{fig3}
\end{figure*}

\begin{table*}[htb]
\scriptsize
\begin{spacing}{1.55}
\begin{ruledtabular}
\begin{tabular}{lccclccc}
\diagbox{error}{percentage}{Class}& Wu &Xiang & Zhu &\diagbox{error}{percentage}{Class}& Wu &Xiang & Zhu \\  
\hline 
& & & &Miscalculation  & 7.6\% & 0 & 0\\
\raisebox{2.1ex}[0pt]{Science error}  & \raisebox{2.1ex}[0pt]{10.2\%} & \raisebox{2.1ex}[0pt]{8.3\%} & \raisebox{2.1ex}[0pt]{14.3\%} & Misusing formulas  & 2.6\% & 8.3\% & 14.3\%  \\
\hline 
& & & &Misunderstanding cycle process  & 15.4\% & 2.1\% & 7.2\%\\
\raisebox{2.1ex}[0pt]{Engineering error}  & \raisebox{2.1ex}[0pt]{71.8\%} & \raisebox{2.1ex}[0pt]{52.1\%} & \raisebox{2.1ex}[0pt]{71.4\%} & Misunderstanding regenerator  & 61.5\% & 50.0\% & 64.2\%  \\
\hline 
No error  & 23.1\% & 45.8\% &14.3\%  \\
\end{tabular}
\end{ruledtabular}
\end{spacing}
\caption{Quantitative results of the errors found in the answers of the three classes lectured by Wu, Xiang and Zhu. The science errors include the errors of miscalculation and misusing formulas, and the engineering errors include the errors of misunderstanding cycle process and misunderstanding regenerator.} 
\label{tab1}
\end{table*}


\subsection{Quantitative description of student difficulties}
The answers given by the students were analyzed case by case. Four types of errors were found in the answers, i.e., miscalculation, The answers given by the students were analyzed case by case. Four types of errors were found in the answers, i.e., miscalculation, misusing formulas, misunderstanding cycle process and misunderstanding regenerator. To get a clear picture of the student performance, we present the percentages of occurrence of the four errors for the three classes lectured by Wu (left panel), Xiang (middle panel) and Zhu (right panel) as well as the case of no error in Fig. 3. Firstly, it is remarkable that more students got correct answer in Xiang class (above 40\%) than the other two classes ($\sim$ 20\%). This could be because the instructor gave the students an engineering tutorial of the heat engine. Secondly, for each class, misunderstanding regenerator is the most common error. One can also notice that, for the error of miscalculation, no one student made this error in classes of Xiang and Zhu, while for Wu class, the bar is divided into three parts. The grey part establishes the percentage of students who only made this error, while magenta part represents the students made the errors of miscalculation and misunderstanding cycle process, simultaneously.  Similarly, the green part shows the percentage of students who made the errors of miscalculation and misunderstanding regenerator, simultaneously. The similar results can be found in the other errors. Generally speaking, the results shown in Fig. \ref{fig3} indicate that most of students adequately understood Stirling cycle process and knew how to do the calculation with the formulas in the textbooks. However, almost all students don't know the function of regenerator. This is due to the pedagogical techniques we used during the teaching.

Since miscalculation and misusing formulas were largely related to the student ability of understanding thermodynamics theory and performing mathematical calculation, they were classified as {\bf science errors}. And since misunderstanding cycle process and misunderstanding regenerator were basically related to the students’ ability to understand the mechanical processes in Stirling cycle, they were classified as {\bf engineering errors}. In fact, we found that many students who misunderstood cycle process also misunderstood regenerator, which make sense since both the errors were caused by the poor comprehensive ability of engineering aspects of Stirling cycle. The results of the quantitative statistics of the errors are listed in Table \ref{tab1}.

The good news is that the science error ratios in the three classes were low, which was 8.3\% (Xiang), 14.3\% (Zhu) and 10.2\% (Wu), respectively. The roughly 10\% error ratio in all the classes shows that the importance of the basic thermodynamics theories was recognized both in the teaching and learning aspects very well and at the approximately same level. This is actually a tradition in the teaching and learning process of most of the physics courses in China.

The bad news is that the engineering error ratios were extremely high, which were above 50\% for all the classes. Detailed analysis show that among the engineering errors, the error of misunderstanding regenerator dominates. For instance, in Xiang class, 52.1\% (25) participants made engineering errors, 50\% (24) participants made the error of misunderstanding regenerator, and 2.1\% (1) participant made the error of misunderstanding cycle process, respectively. The participants who misunderstood regenerator tended to count in the heat absorbed and/or released by the regenerator and obtained the wrong answer for the efficiency. The participant who misunderstood cycle process thought the heat engine did nonzero work during the isovolumic process. Similar results were found in Zhu and Wu classes.

Although the engineering errors were high for the three classes, there was a big difference between Xiang class and the other two classes: the engineering ratio of Xiang class was smaller by approximately 20\% than those of the other two. As we mentioned before, the only difference between the lectures of the three classes was that Xiang class gave a tutorial, which did not cover anything related to Stirling cycle but did introduce general knowledge such as mechanical counterparts of refrigerator corresponding to the reverse Carnot cycle and the meaning of the engineering improvement of heat engines to our society. The tutorial was very preliminary, but it could help some of the students realize that engineering is as important as science in terms of applications of the thermodynamics and really inspire their interest in the engineering aspects of the heat engines. To some extent, the difference between not giving any tutorials and even just giving a preliminary tutorial is like the different between “0” and “1” in computer science. Our results demonstrated how the preliminary tutorial emphasizing engineering aspects of basic knowledges can influence the teaching results.

Another interesting finding is that we can actually use the quantitative results to investigate how the students understand Carnot's theorem. Very few students, for instance, two students in Xiang class and one student in Wu class, made errors in the solution process, but got the right answer for the efficiency of Stirling engine. The written answer sheets or the interviews showed that they used Carnot's theorem to obtain the answer directly after they got stuck in analyzing the cycle processes and performing efficiency calculations. But this usage of Carnot's theorem was kind of passive instead of active, i.e., the students did not realize to use Carnot's theorem until they failed in other ways. In other words, most of the students only correlate Carnot's theorem with Carnot cycle and not with other reversible cycles such as Stirling cycle. The very poor performance on Carnot's theorem was somewhat unexpected, as explicit instruction on it was included in all classes. Our speculation is that this is because for a long time traditional Thermodynamics textbooks only emphasize Carnot cycle and neglect other important cycles. In this sense, it is necessary to emphasize the importance of other cycles besides Carnot cycle. Our results suggest that Stirling cycle could be an important supplement to Carnot cycle.

\section{Summary}\label{sec:summary}
This paper describes an in-depth investigation of student understanding of heat engine in the context of Stirling engine. Our findings indicate that the majority of students in the three classes studied can comprehend the scientific theory of heat engine and make the correct mathematical calculations. However, above 50\% students for all the classes don't understand the basic engineering knowledge of Stirling cycle, especially the function of the regenerator. Importantly, the engineering error ratio of the class that was given a tutorial of engineering knowledge of heat engine was smaller by about 20\% than those of the other two. The results suggest that it is useful to include both scientific and engineering aspects of basic knowledge in instructional approaches, especially in the course of Thermodynamics taught in the countries and regions with a tradition of not paying much attention to experimental teaching or engineering training. In addition, findings from this investigation also suggest that it is necessary to attach more importance to other cycles besides Carnot cycle in the Thermodynamics textbooks to improve the students’ understanding of Carnot’s theorem and related knowledge of heat engine.

\section*{Acknowledgements}
The authors thank the support from New Century Education Reform Project of Sichuan University (Grant No. SCU8147) and College Teaching Research Project of Thermodynamics and Statistical Physics in China (Grant No. JZW-16-RT-07).


\end{document}